\documentclass[prl, twocolumn, superscriptaddress]{revtex4-2}
\usepackage[english]{babel}
\usepackage[pdftex]{graphicx}
\usepackage{color}

\usepackage[sort&compress]{natbib}
\usepackage{hyperref}
\graphicspath{{images/}{./}}

\usepackage{amsmath}
\usepackage{amssymb}
\usepackage{mathtools}
\usepackage{braket}

\renewcommand{\vec}[1]{\ensuremath{\boldsymbol{#1}}}
\newcommand{\un}[1]{\ensuremath{\,\mathrm{#1}}}
\newcommand{\fig}[1]{Figure~\ref{fig:#1}}

\newcommand{\veps}{\varepsilon}
\newcommand{\I}{\mathrm{i}}

\newcommand{\abs}[1]{\left| #1 \right|}
\newcommand{\lr}[1]{\ensuremath{\left( #1 \right)}}

\begin{document}

\title{Atomically thin current pathways in graphene through Kekulé-O engineering}

\author{Santiago Galv\'an y Garc\'ia}
\email{santiagogyg@icf.unam.mx}
\affiliation{Instituto de Ciencias F\'isicas, Universidad Nacional Aut\'onoma de M\'exico, Cuernavaca, M\'exico}

\author{Yonatan Betancur-Ocampo}
\email{ybetancur@fisica.unam.mx}
\affiliation{Instituto de F\'isica, Universidad Nacional Aut\'onoma de M\'exico, Ciudad de México, M\'exico}

\author{Francisco Sánchez-Ochoa}
\email{fsanchez@fisica.unam.mx}
\affiliation{Instituto de F\'isica, Universidad Nacional Aut\'onoma de M\'exico, Ciudad de México, M\'exico}

\author{Thomas Stegmann}
\email{stegmann@icf.unam.mx}
\affiliation{Instituto de Ciencias F\'isicas, Universidad Nacional Aut\'onoma de M\'exico, Cuernavaca, M\'exico}

\date{\today}

\begin{abstract}
We demonstrate that the current flow in graphene can be guided on atomically thin current pathways by means of the engineering of Kekulé-O distortions. A grain boundary in these distortions separates the system into topological distinct regions and induces a ballistic domain-wall state. The state does not depend on the precise orientation of the grain boundary with respect to the graphene sublattice and therefore, permits to guide the current on arbitrary paths through the system. As the state is gapped, the current flow can be switched by electrostatic gates. Our findings can be explained by a generalization of the Jackiw-Rebbi model, where the electrons behave in one region of the system as fermions with an effective complex mass, making the device not only promising for technological applications but also a test-ground for concepts from high-energy physics. An atomic model supported by DFT calculations demonstrates that the proposed system can be realized by decorating graphene with Ti atoms.
\end{abstract}

\maketitle

\section{Introduction}

Controlling and steering the current flow at the nanoscale is an ongoing problem in science and engineering because it is of vast importance for all nanoelectronic devices. One of the most successful strategies has been the electrostatic gating of semiconductors, which allows to switch on and off the current flow and eventually enables devices like field effect transistors which operate in all computer chips. However, further miniaturization of these devices is approaching its end, making it necessary to investigate new ways and materials.

Such a new way to control the current flow is offered by topological insulators, where the bulk is an insulator while the edges feature conducting states which are topologically protected and can be used to guide efficiently the current \cite{Ortmann2015, Asboth2016, Shen2017}. Due to this phenomenal property, topological insulators have become a hot-topic in solid-state research and their discovery has been awarded with the Nobel prize in 2016 \cite{Haldane2017}. In the search of a new material capable of replacing silicon in nanoelectronic devices, graphene will certainly come to mind, given its exceptional transport properties \cite{Neto2009, Katsnelson2012, Torres2020}. However, the absence of a band gap in graphene makes it impossible to switch off the current flow. Even worse, pristine graphene is not topological (as it is gapless) and the Klein tunneling due to the pseudo-spin of the electrons in graphene prevents any confining and guiding of the current flow through electric gates \cite{Katsnelson2006, Young2009, Lee2015, Chen2016}. Several strategies have been discussed how to open a band gap in graphene. For example, narrow nanribbons show a bandgap but are difficult to fabricate although impressive advances have been made by chemical synthesis \cite{Cai2010, Ruffieux2016, Chen2015, Kolmer2020, Wang2021}. Other 2D materials with an intrinsic band gap like the transition metal dichalcogenides (TMDs) or phosphorene suffer from low electron mobility or rapid degradation of the material \cite{Carvalho2016, Manzeli2017}. 

A recently discussed possibility to alter the properties of graphene is through Kekulé distortions where -- inspired by the ideas of August Kekulé for the benzene molecule \cite{Kekule1866} -- the carbon bonds are altered periodically \cite{Gutierrez2015}. It has been shown that graphene on a Cu(111) substrate shows Y-shaped bond alternations and is named therefore Kekulé-Y graphene \cite{Gutierrez2016}. In this case, the Dirac cones are mapped to the $\Gamma$ point and remain gapless but can have different Fermi velocities that can be employed to induce birefringence in a spherical pn junction \cite{Gamayun2018, Andrade2022}. In order to open a gap in Kekulé-Y graphene, it has been proposed using Kekulé-Y bilayers or on-site potentials \cite{Naumis2019, Galvan2022}. Another Kekulé distortion is a benzene-ring-like bond texture named Kekulé-O graphene \cite{Chamon2000, Hou2007, Cheianov2009} that was recently observed in graphene deposited in SiC with Li intercalations \cite{Bao2021}. In this case the electronic structure possesses a band gap proportional to the degree of deformation \cite{Andrade2020}. 

\begin{figure}
    \centering    
    \includegraphics[width=0.95\columnwidth]{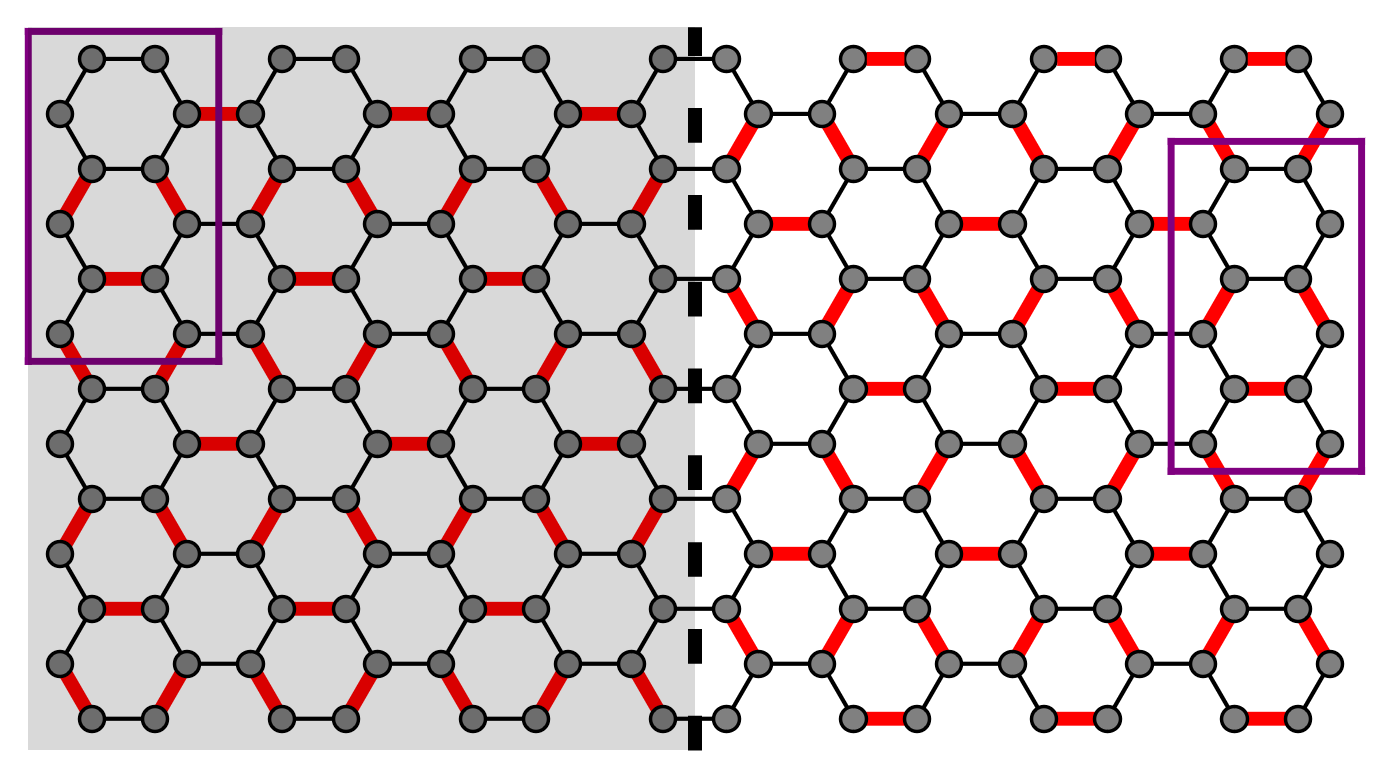}
    \caption{Sketch of the studied system, graphene with Kekulé-O bond distortions where the red coloured bonds are strengthened by $\Delta$. A grain boundary separates the system into two regions, see gray and white background color, the displacement between the atoms in the purple rectangle and the black dashed vertical line.}
    \label{fig:1}
\end{figure}

In this letter, we propose a device, where the current can be guided on arbitrary atomically thin pathways through the system and additionally, switched by electric gates. The current pathways are generated through the engineering of Kekulé-O distortions with a grain boundary, which separate the system into two distinct regions, see \fig{1}. We show that a ballistic domain wall state, named also soliton, arises at the interface of these regions. Our work is a realization (and generalization) of the seminal work by Jackiw and Rebbi \cite{Jackiw1976}, who showed in the context of high-energy physics that a soliton exists at the interface of two topologically distinct regions. Much later the existence of this soliton was demonstrated for the (at that time) newly discovered class of topological insulators \cite{Lee2007}. Semenoff and co-workers suggested the realization of the soliton in a graphene heterojunction composed of two parts with different staggered potentials \cite{Semenoff2008}. More recently, implementations on the basis of polariton graphene \cite{Solnyshkov2022} and even on the basis of Kekulé-O graphene have been proposed \cite{Wu2016, Kariyado2017, Liu2017, Liu2019}. The soliton has not yet been observed directly in graphene but in emulation experiments like photonic crystals \cite{Barik2016, Yang2020} and acoustic resonator networks \cite{Xie2019}. Our paper goes beyond these works in the following way: The soliton arises due to a grain boundary between two regions of Kekulé-O distorted graphene and does not rely on differently modified bonds in the two regions (nor onsite potentials). Most importantly, as realizations of Kekulé-O graphene are still rare, we demonstrate theoretically that our system can be realized by decorations of graphene with Ti atoms. In our case the soliton has a gap and can therefore, be switched efficiently by gates. Moreover, it does not depend on the orientation of the graphene sublattice, allowing to guide the electron current on arbitrary paths by means of suitably engineered grain boundaries. Another more subtle finding is that fact that the electrons in our system behave as a relativistic particles with a \textit{complex} effective mass, making our system not only promising for technological applications but also a test-ground for concepts from high-energy physics. All our findings are supported either by numeric calculations as well as analytical results based on an effective low-energy Hamiltonian. 

\section{System}

We consider a sheet of graphene where certain carbon bonds are changed, see the red bonds in \fig{1}. These changes are in analogy to the Kekulé model for the benzene molecule and therefore, the material is named Kekulé-O (Kek-O) graphene. The system is separated into two regions by a grain boundary, see the vertical displacement of the unit cell framed by the purple rectangle. The position of the grain boundary is indicated by the black dashed vertical line.

The system is modeled by a nearest neighbor tight-binding Hamiltonian, introduced by Gamayun et al. \cite{Gamayun2018}, 
\begin{equation}
    \label{eq:tbh}
    H=-\sum_{\vec{r}} \sum_{j=1}^{3} t_{\vec{r},j}\, a^{\dagger}_{\vec{r}} b_{\vec{r}+\vec{\delta}_{j}}+\text{H.c.}
\end{equation}
where $a^\dagger$ is the creation operator for electrons on graphene's sublattice $\mathcal{A}$ and $b$ the annihilation operator on sublattice $\mathcal{B}$. The vectors $\vec{\delta}_j$ are pointing to the three nearest neighboring carbon atoms. The Kekulé distortions of the carbon bonds are taken into account by 
\begin{equation}
    \frac{t_{\vec{r},j}}{t_0}=1+\frac{\Delta}{3} \Bigl[1{+}2\cos\left[((q{+}1) \vec{K}^+ {+}q \vec{K}^-)\cdot \vec{\delta}_j+\vec{G}\cdot \vec{r}\right]\Bigr],
\end{equation}
where $t_0 \approx 2.8 \un{eV}$ is the unmodified bond with a distance $d_0 \approx 0.142 \un{nm}$ and $\Delta$ its distortion (measured in multiples of $t_0$). The vectors $\vec{K}^{\pm}$ are the high symmetry points and $\vec{G}$ their difference. The parameter $q$ is an integer number with value $1$ and $0$ in the left and right region of \fig{1}, respectively.

\begin{figure*}
    \centering
    \includegraphics[width=2.\columnwidth]{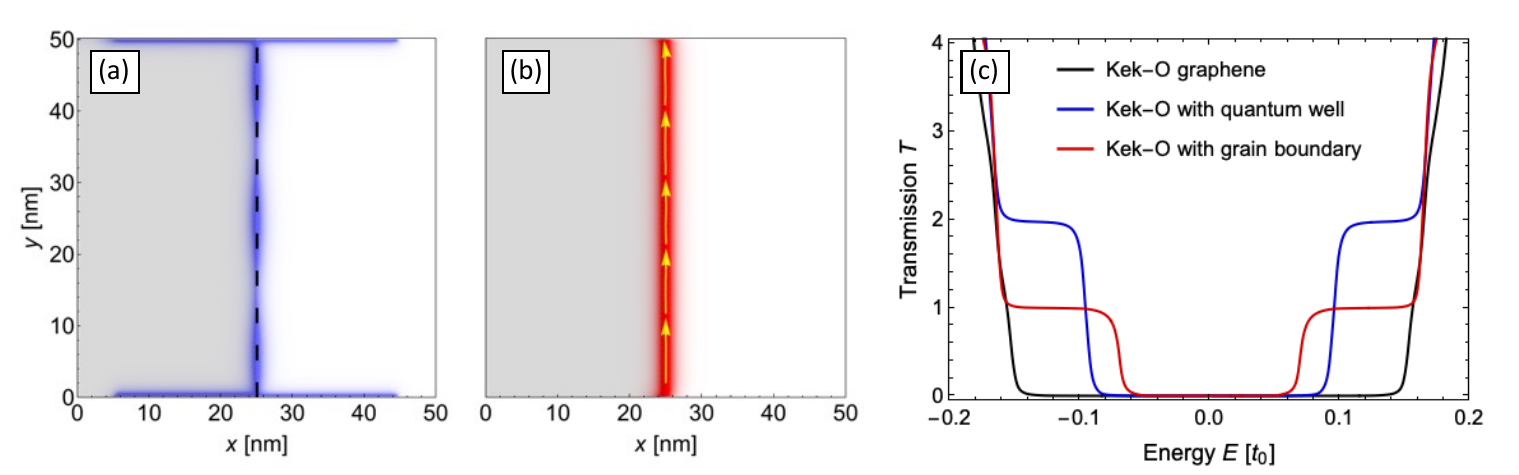}
    \caption{Local density of states (a, blue color shading) and local current (b, red color shading and yellow arrows) for electrons with energy $E=0.5\Delta  210\un{meV}$ injected at the bottom edge. A bond modification of $\Delta=0.15$ is used. A domain wall state is observed clearly at the interface of the two regions (black dashed line). The transmission between the top and bottom contacts (c, red curve) confirms this ballistic state ($T=1$) within the energy range $\Delta/2 \lesssim \abs{E} \lesssim \Delta$. This state is absent in a system without grain boundary (black curve). In the case of a quantum well (blue curve), formed by stripe of pristine graphene between two regions of Kek-O graphene (without grain boundary), we find $T=2$ in a finite energy range.}
    \label{fig:2}
\end{figure*}

Making a Taylor expansion to first order around the $\Gamma$ point, we obtain the effective Hamiltonian of Kek-O graphene
\begin{equation}
 \label{Hk}
    H_\gamma(\vec{p})= v_F
    \begin{pmatrix} 
    \vec{\sigma}\cdot\vec{p}& v_F m_\gamma \sigma_z\\
    v_F m_\gamma^* \sigma_z & \vec{\sigma}\cdot\vec{p}
    \end{pmatrix}
\end{equation}
where the Fermi velocity and the effective electron mass are given by
\begin{equation}
    \label{mq}
        v_F=\frac{t_0  d_0 (3+\Delta)}{2 \hbar},\quad
        m_\gamma= e^{i \gamma} \,\Delta/v_F^2
\end{equation}
with the definition $\gamma= \frac{2\pi}{3}\lr{2q+1}$. We see that $\Delta$ gives a finite mass to the electrons while the parameter $q$ enters in a complex phase. Inserting the two possible values of $q$, we obtain in the left region a positive mass, $m_{\gamma=0}= \Delta/v_F^2$, while in the right region a \textit{complex} effective mass is assigned to the electrons, $m_{\gamma=2\pi/3}= e^{i \frac{2\pi}{3}} \Delta/v_F^2$. Note that the inclusion of a complex mass leaves unaffected the hermiticity in the Hamiltonian and the phase $2\pi/3$ represents the rotation of the distorted benzene rings in the left region with respect to the right region. The energy bands of this Hamiltonian
\begin{equation}
    \label{Ek}
    E(\vec{p})= \pm \sqrt{v_F^2 p^2+\abs{\Delta}^2}
\end{equation}
show the typical dispersion of massive Dirac fermions with a band gap of size $2\abs{\Delta}$. The above energy bands are twofold degenerate due to the valley degree of freedom. They do not depend on the parameter $\gamma$, because for a homogeneous system we can get rid of the complex phase in the mass by a gauge transformation. Note that in other work \cite{Wu2016, Kariyado2017, Liu2017, Liu2019}, different bond modifications are used where in the left region the bonds are strengthened ($t+\Delta$) while in the right region they are weakened ($t-\Delta$), which leads to a negative effective electron mass in the right region and a gapless soliton.

\section{Results}

We investigate the electronic transport in the system and analyze the possibility to steer the current flow on arbitrary atomically thin pathways. For that, further theory to understand our findings is developed. In the following, a bond modification $\Delta=0.15$ is used. In \fig{2}, we show the local density of states (a) and the local current (b) calculated by means of the Green's function method (see the Supplementary Material for details). The electrons have the energy $E=0.5\Delta$ and are injected at the bottom edge. We observe the localization of a domain wall state and a ballistic current at the interface of the grain boundary (dashed black line), which separates the system into two regions (gray and white shaded regions). Note that the LDOS also shows the states induced by the metallic contacts (described by a wideband model) at the bottom and top system edges. \fig{2} (c) displays the transmission between the two contacts, studying in this case a smaller system with semi-infinite leads in order to avoid back-reflections at the system edges. It confirms a ballistic state, $T=1$, within the energy range $\Delta/2 \lesssim \abs{E} \lesssim \Delta$ (red curve). For comparison, we show also the transmission for a system without a grain boundary (black curve), where this ballistic state is absent. The quantization of the transmission at $T=1$ can be understood by the fact that, despite the existence of two degenerate valleys in Kek-O graphene, only a single soliton arises at the grain boundary, see Eq.~\eqref{dredge} and its discussion below. We also added to \fig{2} the transmission for a system consisting of two regions of Kek-O graphene without a grain boundary, which are separated by a small ribbon of pristine graphene (about 8 carbon rings wide), forming effectively a quantum well (blue curve). In this case, we observe a $T=2$ quantization in a certain energy range, which can be distinguished clearly from the $T=1$ for the soliton and is due to a valley-degenerate ballistic state confined in the quantum well. 

\begin{figure}
    \centering
    \includegraphics[width=\columnwidth]{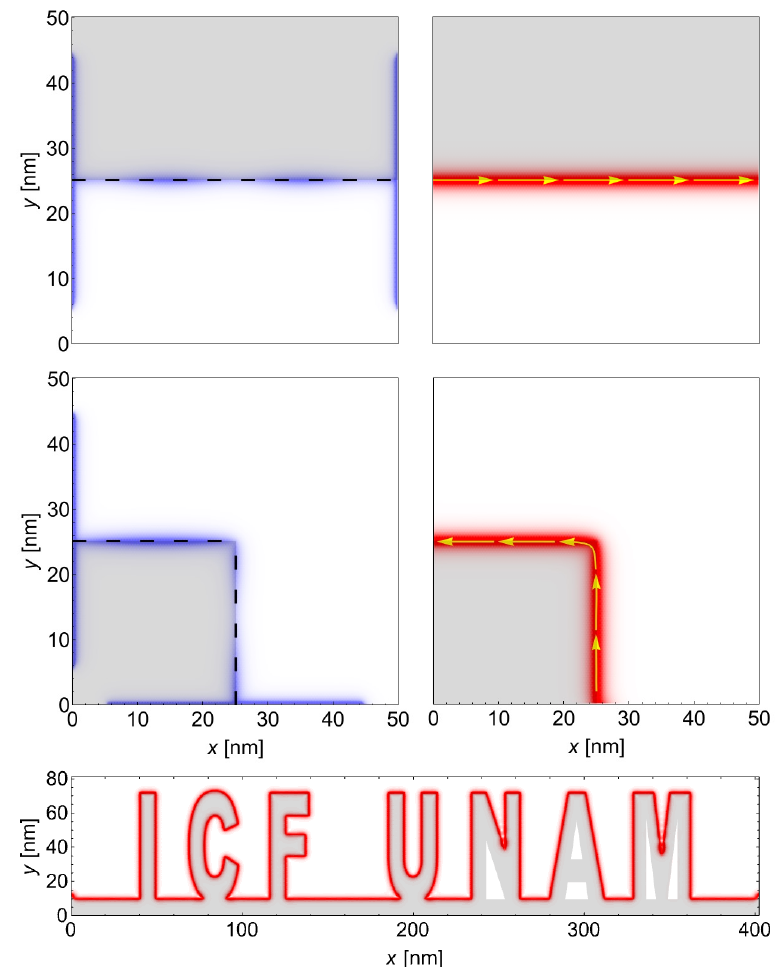}
    \caption{The soliton does not depend on the orientation of the graphene sublattice and therefore, can follow a corner or even more complex patterns, like the initials of our institute.}
    \label{fig:3}
\end{figure}

As shown in \fig{3}, the ballistic state is independent from the orientation of the graphene sublattice with respect to the grain boundary. Even complex current flow patterns can be generated by a suitably shaped grain boundaries, like the initials of our institute. This figure also shows that the precise bond modifications at the interface are not important but the global existence of the grain boundary between the two regions, making them topological distinct. 

\subsection{Generalized Jackiw-Rebbi model}

In the following, we will show that our findings can be explained by a generalized Jackiw-Rebbi model, that was originally developed to explain the emergence of a soliton at the interface between two regions, where in one of them the electrons have a positive effective mass and a negative mass in the other \cite{Jackiw1976}. Here, we generalize this model by permitting a complex valued effective electron mass through the parameter $\gamma$. The model consists in finding the domain wall state through the following Ansatz
\begin{equation}
\begin{split}
    \ket{\Psi_L(x,y)}&=e^{i k_y y}e^{\lambda_L x}\ket{\chi_L},\\
    \ket{\Psi_{R}(x,y)}&=e^{i k_y y}e^{-\lambda_{R} x}\ket{\chi_R},
\end{split}
\end{equation}
where $\ket{\chi_{L/R}}$ are the four-component spinors in the left and right region, respectively. We assume that this state propagates along the $y$ direction and decays exponentially in the $x$ direction. This Ansatz has to fulfill the Schrödinger equations in the two regions
\begin{equation}
\begin{split}
    \left[H_{0}(\vec{p})-E\right]\ket{\Psi_{L}}&=0,\\
    \left[H_{\gamma}(\vec{p})-E\right]\ket{\Psi_R}&=0,
\end{split}
\end{equation}
which allow us to calculate the depth lengths
\begin{equation}
    \lambda_{L/R}=\frac{\sqrt{|m_{0/\gamma}|^2 v_F^4+\hbar^2 k_y^2 v_F^2-E^2}}{\hbar v_F }.
\end{equation}
From the continuity of the wavefunction at the interface, $\ket{\Psi_L (x=0,y)}=\ket{\Psi_R (x=0,y)}$, we obtain finally the energy bands of the soliton
\begin{equation}
    \label{dredge}
    E_\gamma(\vec{p})=\pm \sqrt{p_y^2 v_F^2+\abs{\Delta}^2 \cos^2 (\gamma/2)}.
\end{equation}

\begin{figure}
    \centering
    \includegraphics[width=\columnwidth]{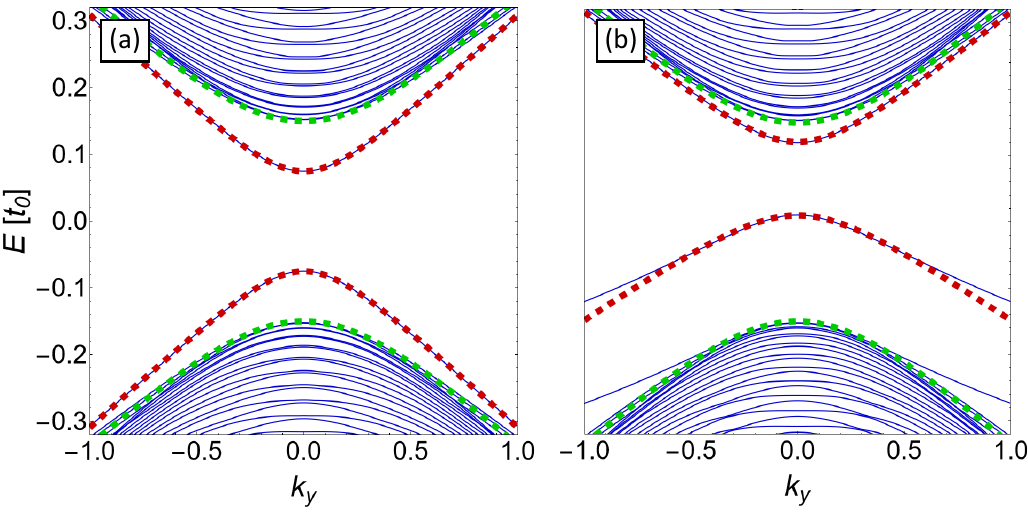}
    \caption{Band structure of the system. (a) The dashed red curve shows the energy band of the soliton from Eq.~\eqref{dredge}, while the green dashed curve shows the energy bands of Kek-O graphene without a grain boundary from Eq.~\eqref{Ek}. The solid blue curves indicate the numerically calculated bandstructure. (b) An onsite potential $\veps=0.3t_0$ at the interface moves the soliton bands to higher energies and reduces the band gap, as observed in the DFT calculations of graphene with Ti adatoms (see \fig{5} below).}
    \label{fig:4}
\end{figure}

In \fig{4} (a) we show this energy band (red dashed curve) together with the energy band from Eq.~\eqref{Ek} for Kek-O graphene without a grain boundary (green dashed curve). The solid blue curves are the numerically calculated energy bands of a finite nanoribbon of Kek-O graphene with grain boundary. Details on the calculations can be found in the Supplementary Material. The analytical models agree perfectly with the numerically calculated energy bands and also match with the transmission plateaus shown in \fig{2}. 

The energy band of the soliton shows a gap of size $\abs{\Delta}^2 \cos^2(\gamma/2)$, which can be modulated through the parameter $\gamma$. In the case of Kek-O graphene with a grain boundary, we have $\gamma=2\pi/3$ and observe that the gap of the soliton $\abs{\Delta}$ is half of the bulk value, in perfect agreement with the transmission curve in \fig{2}. The system of Kek-O graphene where a grain boundary is absent but the bonds in the right region are weakened by $\Delta$ \cite{Wu2016, Kariyado2017, Liu2017, Liu2019}, can be represented by $\gamma=\pi$ and therefore, leads to a gapless edge state.

\subsection{Atomic model \& DFT calculations}

\begin{figure}
    \centering
    \includegraphics[width=0.98\columnwidth]{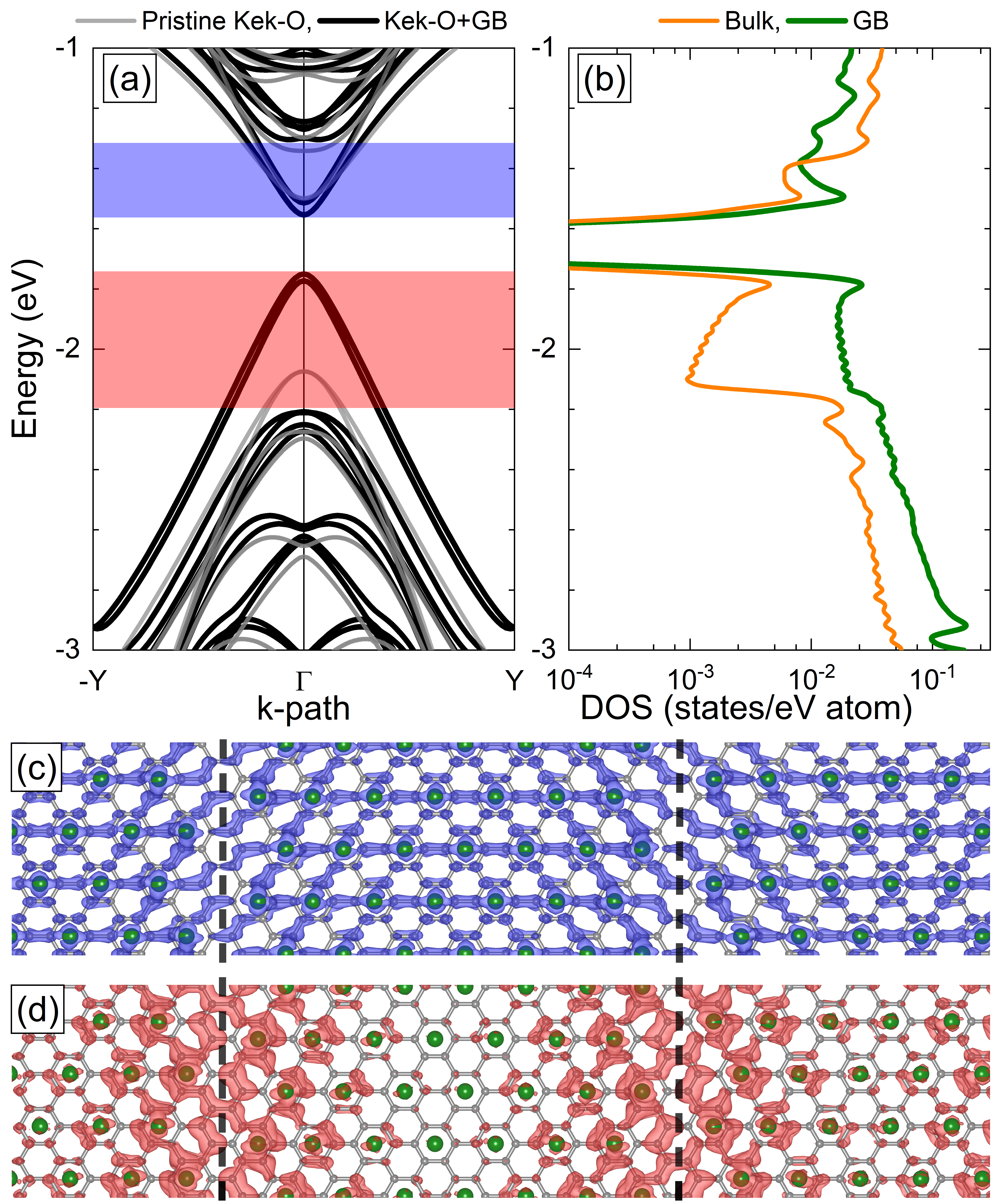}
            \caption{(a) Band structure and (b) average C(2$p_z$) resolved density of states (DOS) for Kekulé-O graphene with grain boundary. The Fermi energy is set to zero eV. (c) and (d) local density of states (LDOS) calculated using the electronic states enclosed by the blue and red shaded regions in (a). Ti and C atoms are in green and gray color, respectively. The vertical dashed black lines denote the atomic grain boundaries.}
    \label{fig:5}
\end{figure}

We propose to realize the system by decorating graphene periodically with Ti atoms on hollow sites of the carbon rings, to perturb the carbon bonds and generate a superlattice with fused Kek-O segments. The atomic model can be seen in \fig{5} (c,d), where the Ti atoms (green spheres) are absorbed on graphene. A grain boundary in the graphene superlattice is generated by a displacement of the Ti atoms. Note that here, we implement two grain boundaries as we use periodic boundary conditions in both, $x$ and $y$ direction. The band structure of this system is calculated by means of DFT (see the Supplementary Material for details) and shown in \fig{5}(a) with black lines, and compared with the band structure of the system without a grain boundary plotted with gray lines. The band gap is reduced considerably by the introduction of a grain boundary, from about $580\un{meV}$ to $200\un{meV}$, in good qualitative agreement with our tight-binding model. In order to analyze the character of the bands close to the gap, we integrate the LDOS over the blue and red shaded energy intervals in \fig{5} (a) and find that the states below the gap are localized at the grain boundary, while the states above the gap can be attributed to a mixing of bulk and grain boundary states. This system property is confirmed by the average resolved DOS for C(2$p_z$) orbitals in \fig{5} (b) which is measured in the bulk (orange curve) and at the grain boundary (green curve), showing that within an energy range of about $500\un{meV}$ below the gap the states are localized predominately at the grain boundary. We can understand this asymmetry by the fact that the Ti atoms will not only change the carbon bonds but also intrinsically dope the system. This doping occurs homogeneously in the bulk but it does not occur at the grain boundary. We can take into account this doping in our tight-binding model by introducing an onsite potential $\varepsilon$ just at the grain boundary. The energy bands in \fig{4} (b) are calculated for such a system with $\varepsilon=0.3 t_0$ and show clearly that the solitonic bands (red dashed curves) are moved upwards to higher energies while the bulk bands (blue curves) remain unchanged. Therefore the upper soliton band mixes with the bulk bands, similar to the DFT calculations. Interestingly, we also observe that the band gap is reduced by the onsite potential and the Fermi velocity of the lower soliton band decreases. Experimentally the atomically precise engineering of Kek-O distortions can be realized by manipulating individual Ti atoms through the tip of a scanning tunneling microscope.

\section{Conclusions}

In this letter, we have shown that the current flow in graphene can be guided on atomically thin pathways by means of the engineering of Kekulé-O distortions. A grain boundary in these distortions (see \fig{1}) separates the system into two regions and generates a soliton at their interface that transports the current ballistically (see \fig{2}). The soliton does not depend on the orientation of the grain boundary with respect to the graphene sublattice and therefore, can be guided on arbitrary paths through the system (see \fig{3}). The soliton is gapped, which permits to switch efficiently the current flow through electrostatic gates. The existence of the soliton can be understood in terms of a generalized Jackiw-Rebbi model, where the electrons have a positive effective mass in one region and a complex one in the other, which makes our system a platform to study phenomena from high-energy physics. Numerically calculated energy bands agree perfectly with the energy bands from a continuous model of Kek-O graphene and the generalized Jackiw-Rebbi model (see \fig{4}). Finally, we have demonstrated by means of DFT calcuations that the proposed system can be realized by decorations of graphene with Ti atoms (see \fig{5}). Technologically, our findings can have important applications in nanoelectronics. In particular, the Kekulé-O engineering in graphene paves the way to nanoelectronic circuits on the scale of individual atoms.

\section{Acknowledgments}
We gratefully acknowledge financial support from UNAM-PAPIIT under Project-ID IN103922, Project-ID IA106223, and CONAHCYT under Project-ID A1-S-13469.

\bibliography{Solit_Kekule.bib}

\pagebreak
\appendix
\section{Supplementary Material:}

\section{Green's function method for electronic transport}

The system properties are studied numerically by means of the Green's function method. As detailed introductions can be found in various textbooks \cite{Datta2005, DiVentra2008}, we summarize here only the essential equations. The Green's function of the system is given by
\begin{equation}
  G(E)= \lr{E-H-\Sigma_S -\Sigma_D}^{-1},
\end{equation}
where $E$ is the energy of the injected electrons and $H$ is the tight-binding Hamiltonian, \eqref{eq:tbh}. The self-energies $\Sigma_{S/D}$ describe the effect of the contacts on the system and are modeled either by semi-infinite leads or the wideband model, which both are physical legitimate. The wideband model represents a generic metallic contact with a constant surface density of states and its self-energy reads
$  \Sigma^\text{wb}=  \sum_{i,j \in \text{contact}} -\I \, t_0 \delta_{ij} \ket{i} \bra{j}$. Semi-infinite leads are capable to model an infinitely extended structure, where back-scattering at the system edges is absent and the self-energy is given by $\Sigma^\text{sinf}=  \tau g^\text{sf} \tau^\dagger$, where $g^\text{sf}$ is the (recursively calculated) surface Green's function of the semi-infinite lead and $\tau$ the matrix which couples it to the central system, see \cite{Lewenkopf2013} for details.

Finally, the current flowing between the atoms at positions $\vec{r}_i$ and $ \vec{r}_j$ is calculated by
\begin{equation}
  I_{ij} = \textrm{Im}(t_{ij}\, (G\, \text{Im}(\Sigma)\, G^{\dagger})_{ij}),
\end{equation}
the local density of states at position $\vec{r}_i$
\begin{equation}
    D_i= \frac{1}{\pi}\lr{G \text{Im}(\Sigma)G^\dagger}_{ii}
\end{equation}
and the transmission
\begin{equation}
    T=4\text{Tr}\lr{G\text{Im}(\Sigma_S)G^\dagger\text{Im}(\Sigma_D)}.
\end{equation}

\section{Tight-binding model for Kek-O graphene nanoribbons}

We consider a Kek-O graphene nanoribbon with zigzag edges, modeled by a nearest-neighbor tight-binding Hamiltonian. Starting with the zigzag edge, the unit cell consists of six atoms, which are linked with the characteristic bonds of the Kekulé-O texture, see \fig{1}. The tight-binding Hamiltonian for this chain is very similar to that one of the Su-Schrieffer-Heeger model, but increased three times, and is given by
\begin{equation}
    H_\textrm{ssh}(k)=\left(\begin{array}{c c c c c c}
        0  & 0 & 0 & t^*(k) & t' (k) & 0  \\
         0  & 0 & 0 & 0 & t^*(k) & t(k) \\
         0  & 0 & 0 & t(k) & 0 & t'^*(k) \\
         t(k) & 0 & t^*(k) & 0 & 0 & 0 \\
         t'^*(k) & t(k) & 0 & 0 & 0 & 0 \\
         0 & t^*(k) &  t'(k) & 0 & 0 & 0
    \end{array}\right),
\end{equation}
where the functions are defined by $t(k) = t e^{i\theta(k)}$, $t'(k) = t' e^{i\theta(k)}$ and $\theta(k) = ka/6$ with the lattice constant $a$. The regular and distorted carbon bonds are denoted by $t$ and $t'$, respectively. For two coupled zigzag chain we obtain the Hamiltonian 
\begin{equation}
    H_\textrm{2ssh}(k)=\left(\begin{array}{cc}
        H_\textrm{ssh}(k) & C \\
         C^\dagger & H'_\textrm{ssh}(k) 
    \end{array}\right),
\end{equation}
where
\begin{equation}
    H'_\textrm{ssh}(k)=\left(\begin{array}{c c c c c c}
        0 & 0 & 0 &  t(k) & 0 & t^*(k)\\
         0  & 0 & 0 & t'^*(k) & t(k) & 0\\
         0  & 0 & 0 & 0 & t^*(k) & t'(k) \\
         t^*(k) & t'(k) & 0 & 0 & 0 & 0 \\
         0 & t^*(k)& t(k) & 0 & 0 & 0 \\
         t(k) & 0 &  t'^*(k) & 0 & 0 & 0
    \end{array}\right).
\end{equation}
is the Hamiltonian of the second chain and the matrix
\begin{equation}\label{C}
    C =\left(\begin{array}{c c c c c c}
        0  & 0 & 0 & 0 & 0 & 0  \\
         0  & 0 & 0 & 0 & 0 & 0 \\
         0  & 0 & 0 & 0 & 0 & 0 \\
         t' & 0 & 0 & 0 & 0 & 0 \\
         0 & t & 0 & 0 & 0 & 0 \\
         0 & 0 & t & 0 & 0 & 0
    \end{array}\right)
\end{equation}
couples them together. 

The tight-binding Hamiltonian of $N$ chains $H_\textrm{Nssh}$ consists of a block tridiagonal matrix, whose size depends on the width of the homogeneous nanoribbon
\begin{align}
\label{eq:Nssh}
    H_\textrm{Nssh}(k) &= \nonumber\\
    &\left(\begin{array}{c c c c c c}
        H_\textrm{ssh}(k)  & C & 0 & 0 & 0 & 0  \\
         C^\dagger & H'_\textrm{ssh}(k) & C' & \cdot & \cdot & \cdot\\
         0  & C'^\dagger  & H_\textrm{ssh}(k) & C & \cdot & \cdot \\
         \cdot & 0 & \cdot & \cdot & \cdot & \cdot \\
         \cdot & \cdot & \cdot & \cdot & \cdot & C \\
         0 & 0 & 0 & 0 & C^\dagger & H'_\textrm{ssh}(k)
    \end{array}\right).
\end{align}
The coupling matrix $C'$ is identical to $C$ in Eq. \eqref{C}, but the sub-diagonal has now the order $\{t,t',t\}$. 

In order to build the Hamiltonian of the system with a grain boundary, as shown in \fig{1}, it is necessary to consider the specific Kek-O texture in the right region, which consists in changing the bonds in Eq. \eqref{eq:Nssh}. The Hamiltonian reads
\begin{equation}\label{Hhet}
    H_\textrm{het}(k) = \left(\begin{array}{c c}
    H^{\textrm{left}}_\textrm{Nssh}(k) & C_0\\
    C^\dagger_0 & H^{\textrm{right}}_\textrm{Nssh}(k)
    \end{array}\right),
\end{equation}
where 
\begin{equation}\label{C0}
    C_0 =\left(\begin{array}{c c c c c c}
        0  & 0 & 0 & t & 0 & 0  \\
         0  & 0 & 0 & 0 & t' & 0 \\
         0  & 0 & 0 & 0 & 0 & t \\
         t' & 0 & 0 & 0 & 0 & 0 \\
         0 & t & 0 & 0 & 0 & 0 \\
         0 & 0 & t & 0 & 0 & 0
    \end{array}\right)
\end{equation}
couples the two regions periodically. Diagonalizing the Hamiltonian $H_\textrm{het}$ in Eq. \eqref{Hhet}, we obtain the electronic band structure shown in \fig{4}.

\subsection{Computational methodology of the atomistic model}
We have built a periodic supercell starting from the rectangular unit cell defined in Fig. \ref{fig:1} with fifteen unit cells along the armchair direction. This is in order to avoid edge states as in graphene nanoribbons and to study two grain boundaries with the topology in Fig. \ref{fig:1} due to periodic conditions. To induce a Kekulé-O distortion, we adsorb Ti atoms on hollow sites of benzene rings. Ti atoms tend to adsorb strongly on hollow sites of graphene monolayer \cite{MANADE2015525}.

DFT calculations were performed within the SIESTA code~\cite{ordejon1996self, soler2002siesta}. The electronic states have been expanded using a linear combination of atomic orbitals (LCAO) with a double-$\zeta$ plus polarized (DZP) basis-set with a PAO.EnergyShift of 50 meV. Here the valence shell for Ti atoms is 4$s$, 3$d$ and 4$p$, while for C atoms is 2$s$, 2$p$ and 3$d$. The exchange-correlation energy has been treated with the Perdew–Burke-Ernzerhof (PBE)~\cite{perdew1996generalized} functional within the generalized gradient approximation (GGA) for solids termed as PBEsol \cite{csonka2009assessing}. Norm-conserving Trouiller–Martins~\cite{troullier1991efficient} pseudopotentials were used to describe core-valence electrons interactions.  A 1$\times$9$\times$1 $k$-grid was used for sampling the reciprocal space with the Monkhorst-Pack scheme~\cite{monkhorst1976special}, and an energy cutoff of 300 Ry for the grid integration of charge density in real space. The electronic temperature was set equal to 0.05 eV with a Methfessel-Paxton statistics. The atomic relaxation was achieved when the inter-atomic forces were $\leq$10 meV/\r{A}, while the electronic relaxation was converged to $10^{-4}$. A vacuum gap of 15 \r{A} in the normal (\textit{z}) direction has been used to prevent interactions between neighbor Ti-doped graphene monolayers in adjacent supercells. Visualization of atomic models and isosurfaces was performed with VESTA program~\cite{momma2011vesta}.

\end{document}